\documentclass[conference]{IEEEtran}
\newcommand{\bl}{\textcolor{black}}
\newcommand{\re}{\textcolor{black}} 
\usepackage{amsmath,amsfonts,amssymb,amsthm}
\usepackage{algorithmic}
\usepackage{array}
\usepackage{subcaption}
\usepackage{multirow}
\usepackage{textcomp}
\usepackage{stfloats}
\usepackage{url}
\usepackage{verbatim}
\usepackage{graphicx}
\usepackage{color, soul}
\usepackage{bm}
\usepackage{cite}
\usepackage{xcolor}
\usepackage{listings}
\usepackage[utf8]{inputenc} 
\usepackage[T1]{fontenc} 
\usepackage{float}
\usepackage{algorithmic}
\usepackage[linesnumbered,ruled]{algorithm2e}

\def\BibTeX{{\rm B\kern-.05em{\sc i\kern-.025em b}\kern-.08em
    T\kern-.1667em\lower.7ex\hbox{E}\kern-.125emX}}
\usepackage{balance}

\begin{document}

\title{QoE-Aware Service Provision for Mobile AR Rendering: An Agent-Driven Approach}

{
    \author{
    \IEEEauthorblockN{Conghao~Zhou\IEEEauthorrefmark{1}, Lulu~Sun\IEEEauthorrefmark{1}, Xiucheng~Wang\IEEEauthorrefmark{1},~Peng~Yang\IEEEauthorrefmark{2},~Feng~Lyu\IEEEauthorrefmark{3}, Sihan Lu\IEEEauthorrefmark{4}, and Xuemin (Sherman) Shen\IEEEauthorrefmark{5}
        \IEEEauthorblockA{
          \IEEEauthorrefmark{1}School~of~Telecommunications Engineering,~Xidian University,~China
        \\\IEEEauthorrefmark{2}School of Electronic Information and Communications, Huazhong University of Science and Technology,~China
        \\\IEEEauthorrefmark{3}School of Computer Science and Engineering, Central South University, China
        \\\IEEEauthorrefmark{4}State Power Investment Corporation Limited,~China
        \\\IEEEauthorrefmark{5}Department~of~Electrical~and~Computer~Engineering,~University~of~Waterloo,~Canada
        \\ conghao.zhou@ieee.org, \{lulusun, xcwang\_1\}@stu.xidian.edu.cn, yangpeng@hust.edu.cn, fenglyu@csu.edu.cn,\\lusihan@spic.com.cn, sshen@uwaterloo.ca}
            }
}

\maketitle

\begin{abstract}
Mobile augmented reality (MAR) is envisioned as a key immersive application in 6G, enabling virtual content rendering aligned with the physical environment through device pose estimation. In this paper, we propose a novel agent-driven communication service provisioning approach for edge-assisted MAR, aiming to reduce communication overhead between MAR devices and the edge server while ensuring the quality of experience (QoE). First, to address the inaccessibility of MAR application-specific information to the network controller, we establish a digital agent powered by large language models (LLMs) on behalf of the MAR service provider, bridging the data and function gap between the MAR service and network domains. Second, to cope with the user-dependent and dynamic nature of data traffic patterns for individual devices, we develop a user-level QoE modeling method that captures the relationship between communication resource demands and perceived user QoE, enabling personalized, agent-driven communication resource management. Trace-driven simulation results demonstrate that the proposed approach outperforms conventional LLM-based QoE-aware service provisioning methods in both user-level QoE modeling accuracy and communication resource efficiency.

\end{abstract}

% \begin{IEEEkeywords}

% \end{IEEEkeywords}

\section{Introduction}

Mobile augmented reality (MAR), a typical form of immersive communication in the future 6G, aims to enrich user perception by overlaying virtual objects onto the physical environment through portable devices, e.g., smart glasses~\cite{zhou2025user}. As a foundational module of MAR, annotation rendering should ensure virtual objects are anchored accurately from the user's perspective based on the real-time estimation of the device pose. Due to the high computation burden of device pose estimation at local MAR devices, the edge-assisted MAR paradigm has become prevalent~\cite{chen2023adaptslam}. Specifically, each MAR device uploads the information on its device poses to an edge server, and the edge server conducts device pose estimation and prepares the corresponding virtual objects for delivering them to the MAR device through communication networks~\cite{8972358}. As a result, future 6G communication networks are envisioned to support high-quality virtual object rendering for providing MAR users with immersive experiences~\cite{shamsabadi2025exploring}. 

However, current communication networks face two key challenges in ensuring MAR user satisfaction through network management. \bl{First, MAR application-specific information typically resides within the over-the-top (OTT) service domain and remains inaccessible to the network controller, creating a fundamental barrier to cross-layer designs for QoE-aware service provisioning.} Specifically, beyond traditional network-related factors, numerous application-specific factors also significantly influence the user experience in MAR. Even under identical network resource allocation policies, users may perceive vastly different levels of immersion since MAR applications adopt diverse operational mechanisms to accommodate varying human behaviors and real-world contexts~\cite{linowes2017augmented}. Therefore, ensuring accurate quality of experience (QoE) for annotation rendering requires the network controller to understand application-specific factors such as device pose estimation mechanisms. Second, data traffic patterns generated by MAR devices are highly personalized, shaped by individual user movements and the surrounding physical environment. Conventional service-oriented resource management techniques in 5G, such as network slicing, fall short in capturing the user-specific quality of service (QoS) requirements when running the same MAR application. This limitation often leads to resource misallocation, ultimately degrading the level of immersion for certain users during annotation rendering~\cite{zhou2024digital,gao2025characterizing,lyu2020lead}. 

To address the aforementioned challenges, recent studies have explored user-level QoS guarantees for MAR~\cite{zhou2024digital}. For instance, the authors in~\cite{zhou2024digital} proposed user-centric service provisioning strategies to meet the delay requirements of individual users for device pose tracking. However, the QoS metrics employed in these studies often fall short of accurately capturing MAR user satisfaction. As a result, many efforts have shifted toward QoE modeling and QoE-oriented communication service provision, yet they largely overlook the unique impact of MAR operational mechanisms on network resource management~\cite{pan2024quality,feng2023qoe}. Our previous work established a digital twin for each individual MAR user, enabling fine-grained modeling of how radio spectrum resource allocation influences the user’s virtual content hit rate, a key QoE metric in MAR~\cite{sun2025qoe}. However, implementing such a framework remains challenging in practice, as the user data used for digital twin establishment is primarily confined to the OTT service domain and is always inaccessible to the network controller due to privacy concerns~\cite{8473381}.

In this paper, we propose an agent-driven approach to facilitate QoE-aware communication service provision for MAR. Specifically, we establish a \emph{digital agent} (DA), acting on behalf of the MAR service provider, by leveraging a large AI model. The DA facilitates interaction between the network controller and the service provider, enabling the integration of data from both the network and MAR service domains. Unlike conventional approaches where the network controller must collect raw application-specific data from the MAR service provider, the DA actively and selectively extracts only the information essential for QoE-aware service provisioning from raw data residing within the MAR service domain, without exposing or transmitting the raw user data itself~\cite{10591707}. 
The main contributions of this paper are as follows:
\begin{itemize}
    \item \bl{We propose an agent-driven service provisioning framework for MAR that bridges the data and functional isolation between the over-the-top MAR service and network domains, thereby facilitating cross-layer design for network resource management.}

    \item We establish an agent on behalf of the MAR service provider using large language models (LLM) and design key service functions of MAR rendering as API tools through model context protocol (MCP), jointly enabling user-specific QoE modeling for MAR.

    \item We develop a QoE-aware communication resource management algorithm \bl{based on the ``tool-calling'' capability of the established agent to handle non-stationary data traffic patterns resulting from unpredictable user movement in MAR.} 
\end{itemize}

\section{System Model and Problem Formulation}

\subsection{Considered Scenario}

As shown in Fig.~\ref{fig:framework}, we consider a multi-user edge-assisted MAR scenario, where users equipped with MAR devices move freely within the communication coverage area of a base station (BS). Each MAR device periodically captures camera frames and renders virtual content into the user’s field of view (FOV) in real time~\cite{campos2021orb}. Denote the set of MAR devices by~$\mathcal{U}$ and the set of camera frames of device~$u \in \mathcal{U}$ by $\mathcal{F}_{u}$. Consider that all MAR devices operate under the same time reference. We use $f$ to denote both the index of a camera frame and the corresponding timestamp at which the frame is captured.

\begin{figure}[t]
    \centering
    \includegraphics[width=0.50\textwidth]{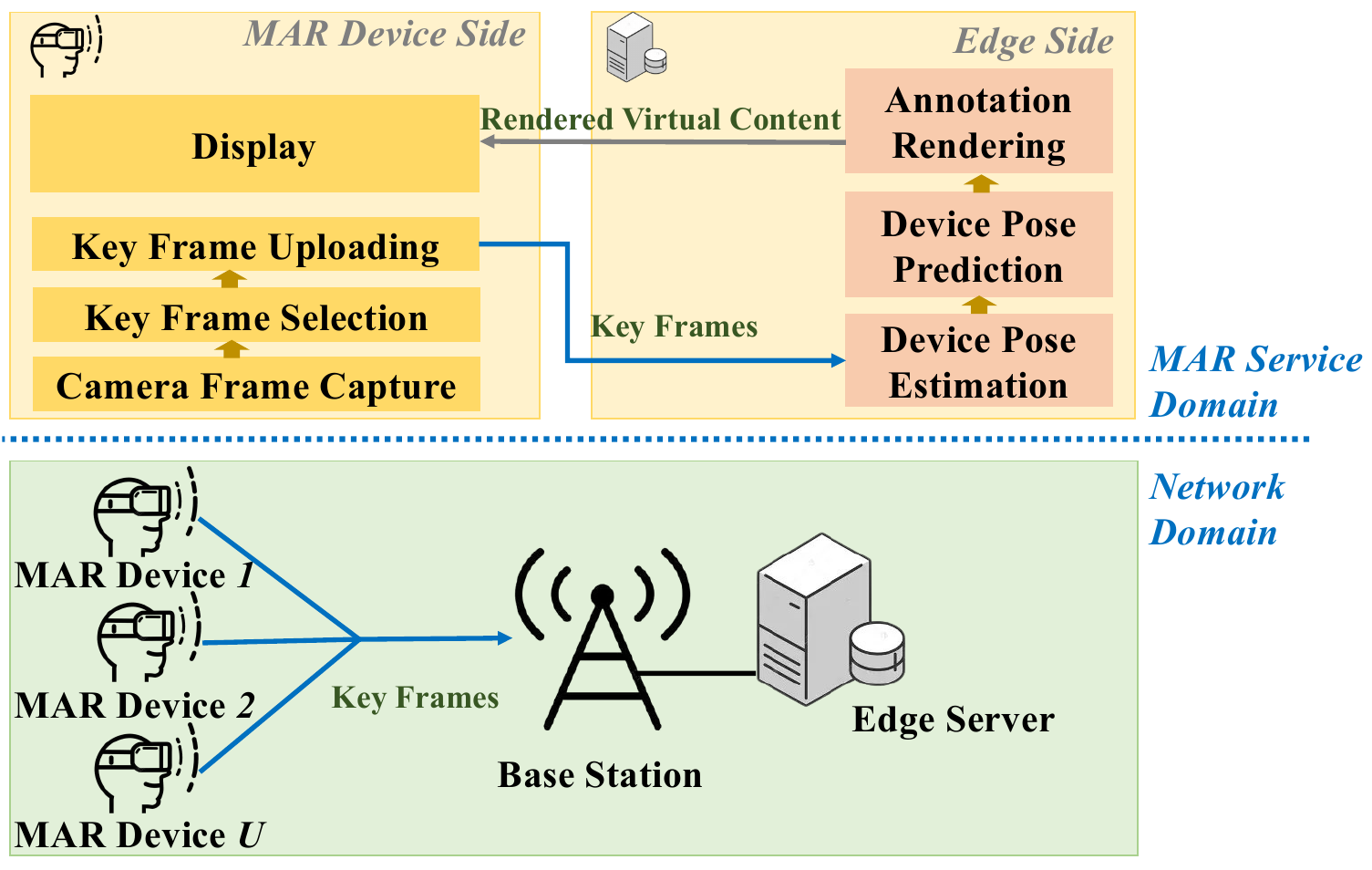}
    \caption{The illustration of the considered scenario.}
    \label{fig:framework}
\end{figure}

The workflow of annotation rendering is summarized as follows~\cite{sun2025qoe}:
\begin{enumerate}
    \item Each MAR device periodically uploads selected camera frames, i.e., key frames, to an edge server co-located with the BS, for device pose prediction. 
    
    \item When a camera frame~$f \in \mathcal{F}_{u}$ is captured, the edge server needs to predict the future 6-DoF pose of the MAR device~$u$ at timestamp~$f+W$, where $W$ is the lookahead window length. For device~$u$, we denote the device pose corresponding to the captured camera frame~$f$ by vector~$\mathbf{q}_{u, f} = [t^{x}_{u,f}, t^{y}_{u,f}, t^{z}_{u,f}, \theta^{x}_{u,f}, \theta^{y}_{u,f}, \theta^{z}_{u,f}]^{\top}$, which characterizes its 3D position and orientation~\cite{chen2023adaptslam}.

    \item The device pose prediction is managed by the MAR service provider, with the actual inference performed by the edge server. Predicting the future device pose relies on a historical sequence of poses, denoted by~$\mathcal{Q}_{u,f} = \{ \mathbf{q}_{u,k} | f-H < k\lambda_{u} \le f \}$. The predicted pose is given by:
    \begin{equation}\label{eq_pose_pred}
        \hat{\mathbf{q}}_{u, f+W} = G_{u}(\mathcal{Q}_{u,f}),
    \end{equation}
    where~$G_u$ denotes the pose prediction function for device~$u$, specific to the MAR service provider. Importantly, the network controller does not always have access to the internal implementation or parameters of $G_u$ during network management.

    \item Using the predicted pose $\hat{\mathbf{q}}_{u, f+W}$, the edge server proactively renders the virtual content anticipated to appear in the user’s FOV, thus facilitating efficient content delivery to the corresponding MAR device.
    
\end{enumerate}

\subsection{Camera Frame Uploading}
As the uplink transmission of camera frames influences the accuracy of device pose prediction, which in turn affects the immersive quality of rendering, it is essential for the network controller to allocate proper spectrum resources to each MAR device to meet the QoE requirements of users. For device~$u$, we model the uplink data rate $r_{u,f}$ at the moment when camera frame $f \in \mathcal{F}_{u}$ is captured as follows:
\begin{equation}
    r_{u,f} = b_{u}\log(1+\gamma_{u,f}),
\end{equation}
where $b_{u}$ represents the spectrum bandwidth (in Hz) allocated to device $u$, and $\gamma_{u,f}$ is the signal-to-noise ratio. 

Due to constraints in uplink communication resources, only a subset of the captured camera frames can be selected for uplink transmission~\cite{chen2023adaptslam}. We assume that the uploaded camera frames are uniformly sampled in temporal order from the set of all captured frames. We define $\lambda_u$ as the uplink sampling frequency (in Hz), i.e., the rate at which user $u$ selects camera frames from the continuously captured camera frame sequence. Consider that the data volume (in bits) of all camera frames for uploading is identical, denoted by $\alpha$. To ensure timely uploading of selected camera frames, the average uplink latency for device~$u$, denoted by $\tau_{u}$, from the moment of frame selection to the completion of transmission, must not exceed a predefined maximum tolerable delay $T$. Based on the D/G/1 queuing model, camera frame uploading should satisfy the following constraint:
    \begin{equation}\label{eq3}
        \tau_{u} \approx \mathbb{E}[S_{u}] + \frac{\lambda_{u}\mathbb{E}[S_{u}^2]}{2(1 - \lambda_{u}\mathbb{E}[S_{u}])} \le T,
    \end{equation}
where $\mathbb{E}[S_{u}]$ denotes the average transmission time for uploading a camera frame from device~$u$, and $\mathbb{E}[S_{u}] = \mathbb{E}[\alpha / r_{u,f}]$. 

\subsection{QoE Model for MAR Rendering}

The perceived level of immersion in MAR is largely determined by how accurately the rendered virtual object aligns with the user’s real-time viewpoint, particularly in terms of spatial placement within the FOV. Therefore, we quantify user QoE at MAR device~$u$ at timestamp~$f$ by virtual content hit rate (VCHR), denoted by $h_{u,f}$. This metric measures the spatial overlap between the virtual object rendered based on the ``predicted'' pose and the content that ``should have been'' rendered based on the user's actual pose at that future moment, given by~\cite{han2020vivo}:
\begin{equation}
    \re{h_{u,f}} = \frac{|\mathcal{C}(\mathbf{q}_{u, f+W}) \cap \mathcal{C}(\hat{\mathbf{q}}_{u, f+W})|}{|\mathcal{C}(\mathbf{q}_{u, f+W}) \cup \mathcal{C}(\hat{\mathbf{q}}_{u, f+W})|},
\end{equation}
where $\mathcal{C}(\mathbf{q})$ denotes the set of virtual object cells visible within the user's view frustum for a given pose $\mathbf{q}$, $\cap$ and $\cup$ denotes the intersection and union of the two sets, and~$|\cdot|$ represents the cardinality of a set. The value of VCHR approaching 1 indicates near-perfect spatial alignment between virtual content and the user’s field of view, reflecting a highly immersive experience. In contrast, a low VCHR value suggests significant misalignment, which can substantially degrade user immersion.

\subsection{Problem Formulation}

Our objective is to reserve sufficient uplink frequency spectrum resources to enhance overall user QoE while reducing spectrum consumption. Define the set of resource reservation decisions as $\mathbf{b} = \{b_u\}_{u \in \mathcal{U}}$ for all MAR devices. The proactive QoE-aware MAR service provision is formulated as follows:

\begin{subequations}\label{p_new}
    \begin{align}
        \textrm{P1:} &\,\, \max_{\mathbf{b}} \sum_{u \in \mathcal{U}} \mathbb{E}[U(h_{u})] - \lambda \sum_{u \in \mathcal{U}} b_{u}\\
        \textrm{s.t.} &\,\, \eqref{eq3} \\
        & \,\, \sum_{u \in \mathcal{U}} b_u \le B_{\text{total}},
    \end{align}
\end{subequations}
where $h_u$ denotes the average QoE for user $u$ based on the values of $h_{u,f}$. Given that the dynamic nature of user behavior renders the future $h_u$ a random variable, our objective is formulated to maximize its expected utility, $\mathbb{E}[U(h_{u})]$, to enable proactive and efficient resource management. where $U(h_{u})$ is a utility function for device~$u$, and $\lambda$ is a parameter for balancing between user QoE and resource consumption, and $B_{\text{total}}$ denotes the total available bandwidth.

Solving Problem~P1 is non-trivial due to two challenges. First, the relationship between allocated resources~$b_u$ and user QoE~$h_u$ is inherently \emph{non-stationary} and \emph{user-specific} since user movement dynamics and environmental contexts evolve over time and differ across users. 
% This means that any fixed, a-priori model mapping resource allocation to QoE will inevitably suffer from performance degradation as the underlying data distribution shifts. 
This compromises the accuracy of estimating $\mathbb{E}[U(h_u)]$ when relying on conventional prediction approaches. Second, the decoupling between network control and the MAR service provider poses a barrier to cross-domain collaboration. For example, the decision making for~$b_u$ lacks access to the information on the pose prediction-related information, i.e.,~$G_u$, which resides within the OTT service domain rather than the network domain. Therefore, enabling QoE-aware service provisioning for MAR requires a novel architecture that \emph{breaks the data and functional isolation between the OTT and network domains}, enabling seamless integrated computation and communication in 6G.

\section{Our Approach}
 In this section, we develop a novel agent-driven approach to address the
two challenges mentioned above.

\subsection{Agent-driven MAR Service Provisioning}
Our key idea is to introduce an agent, acting on behalf of the MAR service provider, that can actively access and analyze user data within the MAR service domain, while exposing or transmitting only the information or insights necessary for QoE-aware service provisioning decision-making, rather than the raw user data itself.

\begin{figure}[t]
    \centering
    \includegraphics[width=0.50\textwidth]{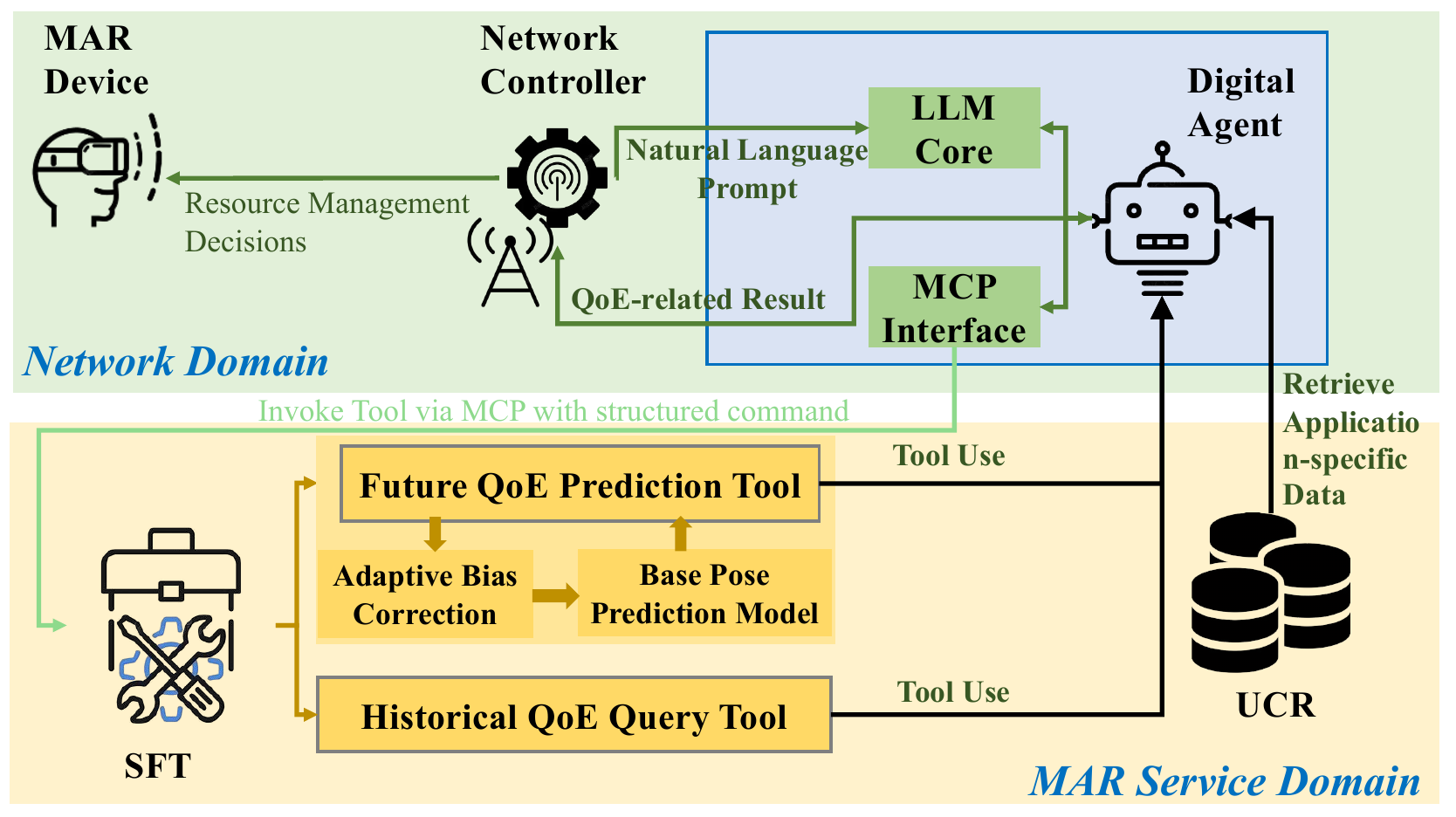}
    \caption{Workflow of the proposed agent-driven service provisioning approach.}
    \label{fig:agent}
\end{figure}

\subsubsection{Cross-Domain Service Provisioning Architecture}

As shown in Fig.~\ref{fig:agent}, we first propose a cross-domain service provisioning architecture, comprising three principal components: i) service function toolkit (SFT); ii) user context repository (UCR); and iii) digital agent (DA). 

\textbf{SFT}: The agent's intelligence is manifested in its powerful tool-using capabilities, rather than in holding or executing any private algorithms itself. This component is designed to abstract key functionalities, e.g., device pose estimation~\cite{campos2021orb}, within the MAR service domain into modular tools accessible by the DA. The interaction between the DA and these tools is based on the Model Context Protocol (MCP). The MCP provides a functional description and a standardized parameter schema for each tool, enabling the DA to intelligently map ambiguous service requests to specific tools. As shown in Fig. 2, our defined key tools in the SFT include: i) Future QoE Prediction Tool encapsulates the deterministic computation of user QoE value. It integrates the service provider's specific base pose prediction model $G_u$ and an adaptive bias correction mechanism using a Kalman filter. By leveraging the tool-using capabilities, the DA can interact with the SFT to conduct accurate and user-specific service demand analysis.
% When the DA invokes this tool with specific user context and network parameters, e.g., a candidate bandwidth allocation $b_u$, it performs complex calculations and returns a precise, deterministic QoE prediction value.
ii) Historical QoE Query Tool enables retrieving and processing a user's historical QoE data, and allowing DA to model the mapping from resource allocation decision~$b_{u}$ and user QoE~$h_{u}$.

\textbf{UCR}: This component is maintained by the MAR service provider to store user data for QoE modeling. During its analysis process, the DA interacts with the UCR. The DA accesses the UCR to extract contextual information needed for user-specific QoE modeling, which is indexed by user ID. When responding to a query, the DA first retrieves the relevant static context from the UCR based on the user ID in the request. Then, it combines this personalized context, e.g., video dataset name and selected rendering algorithm, with dynamic parameters from the network controller, e.g., a candidate bandwidth allocation, to formulate a structured command compliant with the MCP. This command is used to invoke the appropriate tool in the SFT. Ultimately, only the QoE prediction result is returned to the network controller, ensuring that efficient cross-domain collaboration is achieved while protecting data security and model privacy, thereby enabling personalized service provisioning without exposing raw user data to the network domain.

\textbf{DA}: The DA is defined to bridge the MAR service domain and the network domain. Built upon a Large Language Model (LLM) with tool-use capabilities~\cite{ma2024from}, the DA is deployed within the network domain to act on behalf of the MAR service provider. Leveraging the reasoning and abstraction capabilities of the LLM, the DA can invoke service functions such as device pose estimation defined by the MAR provider to perform complex analyses as required. Through carefully crafted system prompt engineering, the DA is capable of interpreting natural language instructions from the network controller using the LLM’s natural language understanding capabilities. For example, given the prompt ``predict user X’s QoE given the allocated 15\,MHz bandwidth'', the DA's first parses the natural language to identify the key entity (user~X) and intent (QoE prediction). It then automatically enriches this by retrieving the user's static service parameters from the UCR and integrating the dynamic network parameters passed directly with the request. Finally, the DA assembles this consolidated information into a standardized, machine-readable command to precisely invoke the QoE prediction tool in the SFT, thus completing the cross-domain query.

\subsubsection{User-Specific and Adaptive QoE Modeling}

To model the user-specific relationship between allocated bandwidth $b_u$ and the resulting user QoE, the proposed cross-domain framework leverages the SFT, where key service functionalities from the MAR domain are encapsulated into modular, API-callable tools. These tools can be directly invoked by the DA, allowing it to compute QoE values without requiring the network controller to access or build mathematical QoE models, i.e.,~$G_u$. In particular, the DA can utilize user-specific context retrieved from the UCR to support personalized QoE prediction, enabling QoE-aware service provisioning.

To handle the non-stationary relationship between bandwidth and QoE value of an individual user, an adaptive bias correction mechanism using a Kalman filter is proposed. This mechanism treats the prediction error $e_t = q_t - \hat{q}_{corrected,t}$ as a real-time measurement of a time-varying bias $b_t$. The filter continuously updates its estimate of this bias $\hat{b}_{t}$ and applies it to correct subsequent base predictions:
\begin{equation}
    \hat{q}_{corrected,t+W} = \hat{q}_{base,t+W} + \hat{b}_{t}.
\end{equation}
This closed-loop process enables the model to track and compensate for drifts in user behavior, ensuring prediction accuracy in dynamic environments.

\subsection{QoE-aware Service Provisioning}

Given predicted QoE values provided by the DA, we propose an agent-driven service provision algorithm summarized in Algorithm~1 to solve Problem~P1. Specifically, two classes of users can be identified: bandwidth ``donors'', whose predicted QoE exceeds a predefined high threshold, $h_{hig}$, and ``receivers'', whose predicted QoE falls below a target threshold, $h_{tar}$. Subsequently, the surplus bandwidth, $B_{sur}$, recovered from the donors is proportionally distributed among all receivers. Specifically, the additional bandwidth, $\Delta b_{u'}^{+}$, allocated to each receiver $u'$ is determined by the magnitude of their QoE deficit, as follows:
\begin{equation}
    \Delta b_{u'}^{+} \leftarrow B_{sur} \cdot \frac{h_{tar} - \hat{h}_{u'}}{D_{all}},
\end{equation}
where $D_{all}$ represents the total QoE deficit of all receivers. This mechanism ensures that system resources are fluidly directed to where they are most needed, thereby enhancing overall spectrum efficiency while preserving user fairness.

\begin{algorithm}[t]
\caption{Agent-driven Service Provision  Algorithm}\label{alg:resource_alloc}
\KwIn{$\mathcal{U}$, $\{b_u\}$, $h_{\text{tar}}, h_{\text{hig}}$}
Initialize $B_{\text{sur}}\leftarrow 0$, $D_{\text{all}} \leftarrow 0$, Donors $\leftarrow \emptyset$, Receivers $\leftarrow \emptyset$\;
\ForEach{user $u \in \mathcal{U}$}{
    $\hat{h}_u \leftarrow \text{PredictFutureQoE}(u, b_u)$\;
    \If{$\hat{h}_u > h_{\text{hig}}$}{
        Find $b_u' < b_u$ such that $\text{PredictFutureQoE}(u, b_u') \approx h_{\text{tar}}$\;
        $\Delta b_u^{-} \leftarrow b_u - b_u'$, $B_{\text{sur}} \leftarrow B_{\text{sur}} + \Delta b_u^{-}$\;
        $b_u^{\text{new}} \leftarrow b_u'$, Add $u$ to Donors\;
    }
    \ElseIf{$\hat{h}_u < h_{\text{tar}}$}{
        Add $u$ to Receivers\;
        $D_{\text{all}} \leftarrow D_{\text{all}} + (h_{\text{tar}} - \hat{h}_u)$, $b_u^{\text{new}} \leftarrow b_u$\;
    }

}
\If{$D_{\text{all}} > 0$ and $B_{\text{sur}} > 0$}{
    \ForEach{user $u' \in \text{Receivers}$}{
        $\hat{h}_{u'} \leftarrow \text{PredictFutureQoE}(u', b_{u'})$\;
        $\Delta b_{u'}^{+} \leftarrow B_{\text{sur}} \cdot \frac{h_{\text{tar}} - \hat{h}_{u'}}{D_{\text{all}}}$\;
        $b_{u'}^{\text{new}} \leftarrow b_{u'}^{\text{new}} + \Delta b_{u'}^{+}$\;
    }
}
\KwOut{$\{b_u^{\text{new}}\}$}
\end{algorithm}

\section{Performance Evaluation}

\bl{In this section, we conduct trace-driven simulations to validate the efficacy of our proposed agent-driven approach in user-specific QoE modeling and QoE-aware communication resource management.} 

\subsection{Simulation Settings}
We investigate a scenario involving 40 MAR users. To realistically simulate individual user movement, we utilize a real-world 6DoF pose dataset (\url{https://github.com/Yong-Chen94/6DoF_Video_FoV_Dataset}). This dataset contains the head movement trajectories of 40 subjects as they viewed a statically rendered volumetric video titled ``Longdres''. The video sequence is presented at 30 frames per second and has a duration of 10 seconds. In our experimental setup, the virtual content of each frame is structured as a point cloud, which is spatially partitioned into a $4 \times 4 \times 2$ grid of cells. Following the methodology in~\cite{han2020vivo}, we determine the visible cell set based on the user's ground-truth and predicted poses, defined as the non-occluded cells residing within the viewing frustum. We adopt the following two AI-driven benchmarks for performance comparsion: 
\begin{itemize}
        \item \textbf{Base Transformer (BT)}: A standard Transformer model trained on aggregated data from all users.
        
        \item \textbf{Personalized Transformer (PT)}: \bl{A Transformer-based model is trained separately for each MAR user category using its corresponding historical data. Additionally, an auxiliary classifier (NETLLM) is trained~\cite{10.1145/3651890.3672268} to identify or assign a user category based on individual user data.}
\end{itemize}

\subsection{Performance of QoE Modeling}

We employ two primary metrics to evaluate the performance of QoE modeling. First, we use the mean squared error (MSE) between the predicted and actual QoE values. Second, we define ``User Category Accuracy'' as the proportion of correctly classified VCHR intervals. Specifically, the continuous VCHR values, i.e.~$h_{u,f}$, in the range $[0.0, 1.0]$ are uniformly discretized into 10 equal-width bins, and a prediction is considered correct if it falls into the same bin as the corresponding ground-truth value.

\begin{figure}[t]
    \centering
    \includegraphics[width=0.40\textwidth]{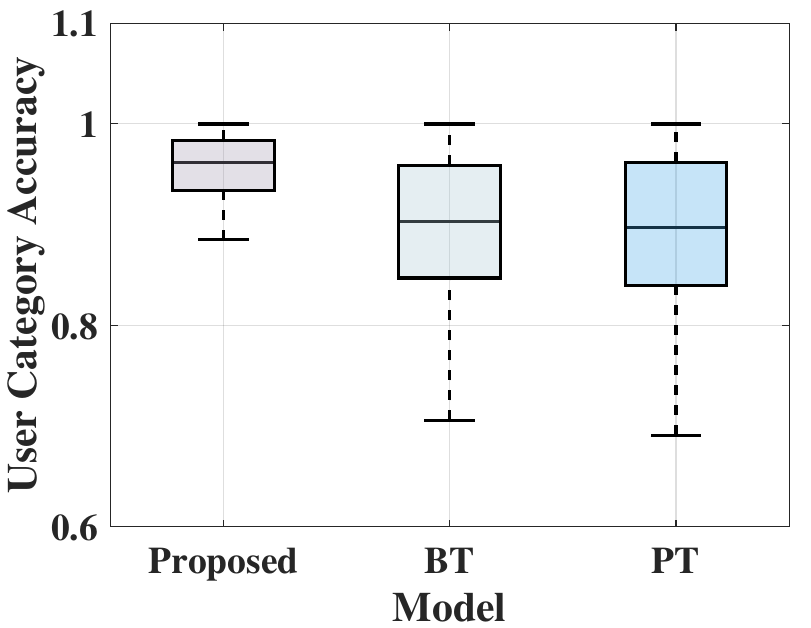}
    \caption{The comparison of user category accuracy.}
    \label{fig:fig4}
\end{figure}
\begin{figure}[t]
    \centering
    \includegraphics[width=0.40\textwidth]{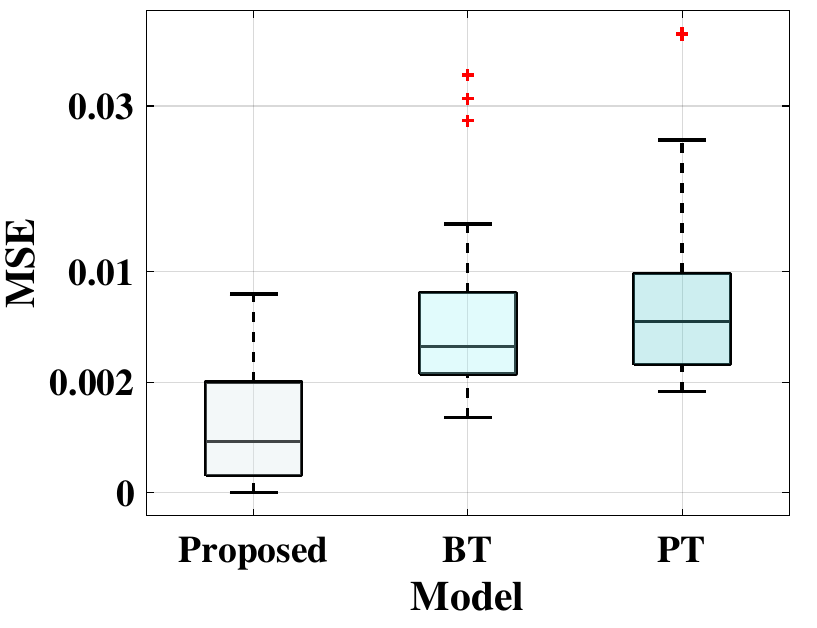}
    \caption{The comparison of MSE for user-level QoE modeling.}
    \label{fig:MSE}
\end{figure}

As shown in Figs.~\ref{fig:fig4} and~\ref{fig:MSE}, we compare our proposed agent-driven approach (labeled as ``PM'') with the two benchmark approaches in terms of ``MSE'' and ``User Category Accuracy'', respectively. We observe that the agent-driven approach outperforms both benchmark methods across the two evaluation metrics. The ``Personalized Transformer'' achieves strong performance when the test data for a user closely resembles their training data. However, its limited training samples may lead to a sharp performance drop when confronted with new or previously unseen user behaviors. In contrast, the ``Base Transformer'' demonstrates robustness to newly emerging data patterns, as it is trained on aggregated data from all users, effectively capturing user-averaged behaviors. Nonetheless, this generalization comes at the cost of precision, preventing it from making highly accurate QoE modeling for any individual user.

Our agent-driven approach effectively circumvents these limitations due to two primary advantages. First, the SA can deterministically calculate the QoE value for an individual user by leveraging application-specific information, such as MAR operational mechanisms, through access to the UCR and invocation of MAR service functions via the MCP. As a result, the performance of QoE modeling does not rely on approximating user-specific QoE based on historical data. This deterministic calculation ensures high precision and reliability in user-level QoE modeling, fundamentally distinguishing our approach from deep neural networks (DNNs), which attempt to approximate a QoE model from history user data.

Second, the proposed agent-driven approach enables user-specific QoE modeling to capture non-stationary data traffic patterns at the individual user level. By integrating a Kalman filter, the approach allows the QoE model to continuously track and correct prediction bias caused by dynamic user behavior, using real-time error observations. This adaptive capability is critical and stands in contrast to static, pre-trained DNNs, which lack the flexibility to respond to such temporal variations.

\subsection{Performance of QoE-aware Service Provisioning}
\begin{figure}[t]
    \centering
    \includegraphics[width=0.37\textwidth]{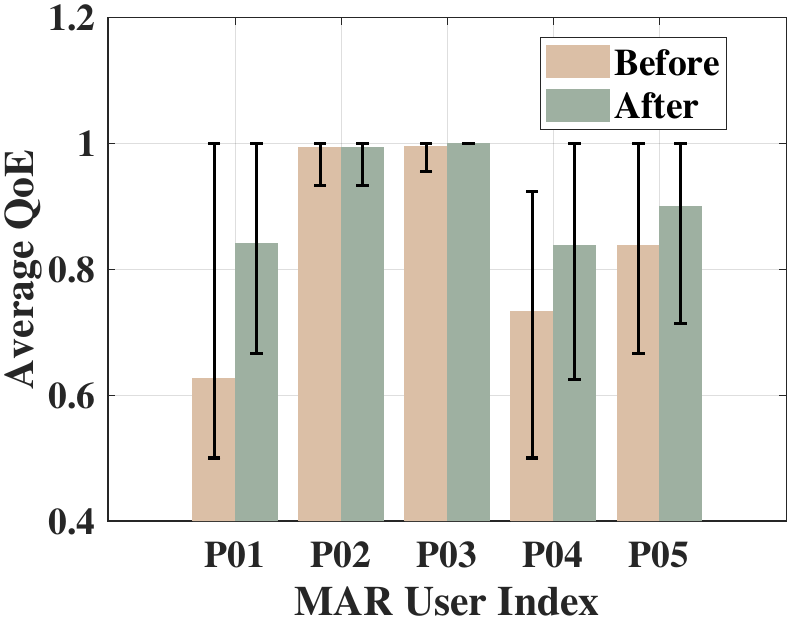}
    \caption{\bl{QoE values before and after agent-driven service provisioning}}
    \label{fig:zhuzhuangtu}
\end{figure}
To validate the efficacy of our agent-driven approach in communication resource management, we consider a scenario involving five users. As shown in Fig~\ref{fig:zhuzhuangtu}, the chart illustrates the average QoE for each user under two distinct resource allocation strategies. The dark brown bars and dark green bars represent the average QoE results before and after the decision-making on service provisioning, respectively. The DA is able to leverage its tool-calling capabilities to proactively identify that Users P01, P04, and P05 would experience significant QoE degradation under the uniform bandwidth allocation. To mitigate this issue, the DA enables the network controller to proactively reallocate surplus bandwidth from Users P02 and P03, who exhibited high initial QoE, to these three users. The resulting improvements are reflected in the dark green bars. The average QoE for Users P01, P04, and P05 is significantly enhanced, while the QoE values of Users P02 and P03 remain at a high level.

\section{Conclusion and Future Work}

In this paper, we have proposed a novel framework for QoE-aware service provisioning in MAR. Based on large language models, we have established a service agent for the MAR service provider to enable collaboration between the network and MAR service domains while avoiding the exposure of MAR user data to the network domain. Second, we have defined and encapsulated key MAR service functions as tools, allowing the service agent to capture user-specific resource demands given QoE requirements. Third, we have proposed a QoE-aware spectrum bandwidth management approach to enhance user immersion during MAR rendering. Our framework for cross-domain collaboration can be extended to various 6G services that require intensive application-specific information for resource provisioning. In the future, we will investigate intent-driven collaboration between the network controller and the service agent based on the agentic AI technique.

\bibliography{references}

\bibliographystyle{IEEEtran}

% \bibliographystyle{IEEEtran}
% \begin{thebibliography}{1}

% \end{thebibliography}

\end{document}